\documentclass[pra,twocolumn,superscriptaddress,showpacs]{revtex4}

\usepackage{amsbsy}
\usepackage{amsfonts}
\usepackage{amssymb}
\usepackage{amsmath}
\usepackage{graphicx,color}
\usepackage{graphicx}

\newcommand{\ket}[1]{| #1 \rangle}

\newcommand{\ie}{{\it{i.e.~}}}
\newcommand{\etal}{{\it{et al.}}}

\definecolor{gray}{rgb}{.4,.4,.4}
\definecolor{deepgreen}{rgb}{.1,.6,.3}

\begin{document}


\title{Protection of quantum information and optimal singlet conversion through higher dimensional quantum systems and environment monitoring }

\author{E. Mascarenhas}
\author{B. Marques}
\affiliation{Departamento de F\'isica, Universidade Federal de Minas Gerais, Belo Horizonte, Caixa Postal 702, 30123-970, MG, Brazil}
\author{D. Cavalcanti}
\affiliation{ICFO-Institut de Ciencies Fotoniques, Mediterranean
Technology Park, 08860 Castelldefels (Barcelona), Spain}
\affiliation{Center for Quantum Technologies, University of Singapore, Singapore}
\author{M. Terra Cunha}
\affiliation{Departamento de Matem\'atica, Universidade Federal de Minas Gerais, Belo Horizonte, Caixa Postal 702, 30123-970, MG, Brazil}
\author{M. Fran\c{c}a Santos}
\affiliation{Departamento de F\'isica, Universidade Federal de Minas Gerais, Belo Horizonte, Caixa Postal 702, 30123-970, MG, Brazil}

\begin{abstract}
We study how to protect quantum information 
in quantum systems 
subjected to local dissipation. We show that combining the use of three-level systems, environment monitoring, and local feedback
can fully and deterministically protect any available quantum information, including entanglement initially shared by different parties. These results can represent a gain in resources and/or distances in quantum communication protocols such as quantum repeaters and teleportation as well as time for quantum memories. Finally, we show that monitoring local environments physically implements the optimum singlet conversion protocol, essential for classical entanglement percolation.
\end{abstract}
\maketitle

\section{Introduction}
Quantum information processing relies on the
capacity of sustaining coherence as well as the entanglement among different
parties. However, many systems proposed for implementing such protocols are naturally subjected to local
and unavoidable dissipation such as ion traps \cite{review ion}, cavity QED systems \cite{QED}, and atomic ensembles~\cite{atomicE}. Needless to say, this  usually works against their quantum efficiency through the
mechanism of decoherence. Different strategies 
have been designed to partially protect or restore this
coherence, such as entanglement distillation~\cite{Distill}, quantum repeaters~\cite{Repeat} 
or feedback~\cite{Feedback}, even though all these proposals present combinations of different costs like the increase of 
the amount of resources to create the necessary redundancy or the need to perform
multi-qubit, \textit{i.e.} non-local, operations. 

In this paper we show a fully deterministic and \emph{local} scheme that counteracts the unavoidable action of dissipation. It relies on three basic elements: 
the capacity to monitor local environments detecting whether one or none excitation
has leaked into each local reservoir; 
the possibility to rapidly feedback the excitation 
into the lossy qubit; and the use of three-level systems to encode qubits.
Note that even though the used encoding implies some sort of redundancy, the scheme is much less demanding than 
the usual quantum error correction codes for dissipative systems~\cite{Feedback} because it adds only one local extra level to each part 
and does not require any extra entanglement or global operation. We discuss applications of this idea to improve the
efficiency of quantum repeater protocols and to produce longer lasting coherence and/or
entanglement for quantum information storage, teleportation or swapping. We also show that it can represent
a physical implementation of the optimum singlet conversion protocol~\cite{Vidal}.

Let us consider a global system composed of internal subsystems that are weakly coupled to
their own local reservoirs. These couplings should respect typical
Markov and Born approximations and one should
be able to read information from each environment regarding the emission of
single excitations from the respective subsystem in a time scale much shorter
than the subsystem decay rate and still much larger
than the correlation times of its respective reservoir. We then show that, within the
framework of the well known quantum trajectories technique~\cite{RevTrajec}, 
the monitoring of the local environments plus the classical communication of the obtained results (hence,
the first element of the scheme) is already enough to enhance the quantum communication efficiency. 
However, in the common case of qubits encoded in two-level systems, 
only the so-called no-jump trajectory, obtained  when
no environment is disturbed, is of use. In fact, the scheme starts to fail 
as soon as any single qubit decays, since its particular state is projected 
to the low energy level and factored out of the rest of the system. In order to solve
this problem, we introduce the two remaining elements to show that encoding the qubits in three-level systems,
combined with feedback on individual systems powers up the scheme to the limiting point 
of completely suppressing the action of the local environments only through local operations.

\section{Monitoring the environment}
\begin{figure*}
\centering
\includegraphics[scale=1]{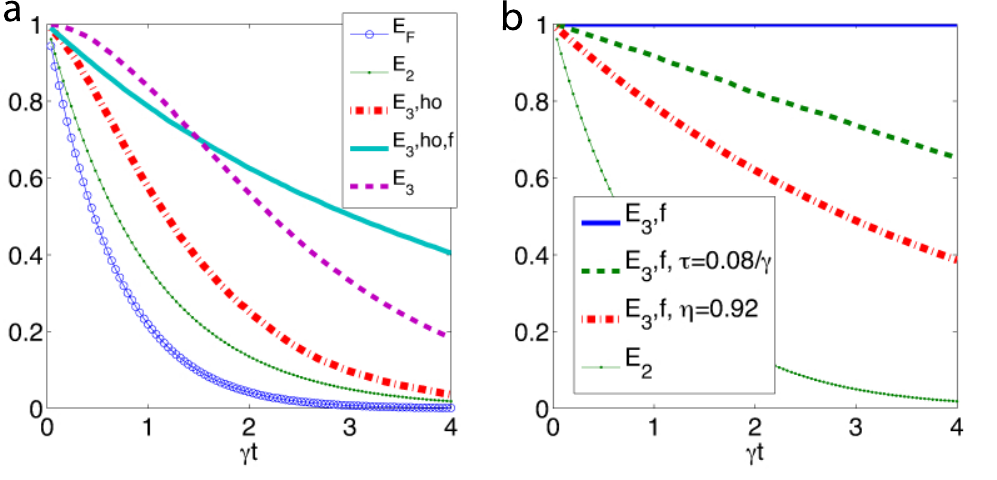}
\caption{(Color online)
Average entanglement over different trajectories, given by $E=\sum_{j_i}E(\psi^{j_i})P^{j_i}$, as a function of $\gamma t$. $E_F$ represents the entanglement evolution without any protection strategy. The numerical index corresponds to the dimension of the system in which both logical qubits are encoded. Index ``f'' indicates the presence of feedback in the system. In $E_{3}$ and $E_{3,f}$ $\gamma_{21} =\gamma_{10}$ (fully degenerated decay channels) whereas index ``h.o.'' indicates decay channels similar to the dissipative harmonic oscillator, i.e. $\gamma_{21} = 2\gamma_{10}$. $\tau$ is the feedback time delay, and $\eta$ is the measurement efficiency. In the right panel we repeat the $E_2$ curve to emphasize that the use of qutrits is advantageous even in the cases with non-perfect measurements or feedback.}\label{fig1}
\end{figure*}

To start with, let us take a
look at how environment-monitoring can enhance a simple quantum
teleportation scheme for which Alice and Bob need to
share a pair of maximally entangled qubits. Consider a initial
Bell state given by $|\Psi_0\rangle = \frac{|10\rangle +
|01\rangle}{\sqrt{2}}$ and that each qubit is under the action of
a local spontaneous emission reservoir. Without external
monitoring, the joint state will become a mixture and the
pre-existing entanglement will eventually vanish. The time
evolution of the whole system is described by the following master
equation:
\begin{eqnarray}
\frac{d \rho}{dt} = -\frac{\gamma}{2} (\hat{a}^\dagger \hat{a}
\rho + \rho \hat{a}^\dagger \hat{a}) + \gamma \hat{a} \rho
\hat{a}^\dagger
 -\frac{\gamma}{2} (\hat{b}^\dagger \hat{b} \rho + \rho \hat{b}^\dagger \hat{b}) + \gamma \hat{b} \rho \hat{b}^\dagger,
 \nonumber
\end{eqnarray}
where we have considered $\gamma_{A}=\gamma_{B}=\gamma$ as the
local decay rates, $\hat{a}$ and $\hat{b}$ are the lowering
operators for the qubits of Alice and Bob respectively, and the
interaction picture is implied. As time goes by, 
entanglement is lost, which can be evidenced by the
entanglement of formation $E_F$ of state $\rho(t)$ ~\cite{EfC} as shown in Fig. \ref{fig1}. 
Due to this process, the further
Alice and Bob take to perform the teleportation, the largest the
amount of resources needed, since they would need more and more
copies in order to distill a maximally entangled pair.

Let us now assume that Alice and Bob can monitor the environments
of their respective qubits, \ie they can continuously detect if
their qubits have lost one excitation. It can involve 
a direct measurement of the reservoir, for example, by detecting an
emitted photon, or an indirect probing of the loss of excitation through an 
auxiliary system~\cite{HarocheNature}. Under these conditions, the time
evolution of the system is no longer described by the master
equation itself but rather by its so-called quantum-jump
unraveling, in which the no-jump operator (corresponding to no detection in the monitored environment) 
is given by $\hat{\Pi}_{0} = \openone-\frac{dt\gamma}{2}[\hat{a}^\dagger \hat{a}+\hat{b}^\dagger \hat{b}]$ 
and the one-jump operator is given by $\hat{\Pi}_{1,A}=\sqrt{dt
\gamma} \hat{a}$~\cite{RevTrajec}. The same holds for Bob.

First, since a detection in the
environment immediately kills the entanglement, 
only the no-jump trajectory is useful. Furthermore, given that 
$\gamma_A=\gamma_B$, the initial state
is preserved under this same evolution, \ie
$|\Psi(kdt)^{NJ}\rangle=|\Psi(0)\rangle$ for any $k$, keeping its
original entanglement. The effect of the reservoir is to make the
no-jump trajectory less and less probable with time, its
probability given by $P_{NJ}=e^{-\gamma T}$. Provided that
Alice and Bob can classically communicate the absence of 
jumps in their environments, this method is an alternative to 
the usual distillation protocol since at any time the two parts
share a maximally entangled state. 


\subsection{Optimal singlet conversion through environment monitoring}

The idea of locally monitoring the reservoir can also be used in the problem of converting a partially entangled state $\ket{\psi_0}=\sqrt{\alpha}\ket{00}+\sqrt{1-\alpha}\ket{11}$
(with $0\leq\alpha\leq 1/2$) into a maximally entangled one (singlet conversion) \cite{Vidal}. Following a similar reasoning to the 2-qubit protocol presented before, when the subsystems undergo spontaneous decay and no jump is detected in both local reservoirs the
initial state $\ket{\psi_0}$  evolves into
$\ket{\tilde{\psi}'}=\sqrt{\alpha}\ket{00}+\sqrt{1-\alpha}e^{-\gamma
t}\ket{11}$, with probability $\alpha+(1-\alpha)e^{-2\gamma
t}$.
When $e^{-\gamma
t}=\sqrt{\alpha/(1-\alpha)}$, $\ket{\psi'}$ is a maximally entangled
state~\cite{Alejo}. This happens with probability $p_{ok}=2\alpha$, which is exactly the optimal singlet conversion probability found in
\cite{Vidal}. This shows that indeed coupling the qubits with
independent reservoirs and monitoring their environments provides a physical implementation of the
optimal singlet conversion  protocol. A setup that implements this conversion has recently been done with twin photons \cite{Alejo}.

\section{Encoding qubits in three-level systems}

We now proceed to show that  using 
qutrits ($\{|0\rangle,|1\rangle,|2\rangle\}$) instead of qubits in
each side improves the previous strategies. Alice and Bob will always share an initial state in the subspace spanned by $\{|1\rangle,|2\rangle\}_A\otimes \{|1\rangle,|2\rangle\}_B$, which will
represent our particular encoding. We will also consider, from now on, only cascade decay channels, 
$|2\rangle \mapsto |1\rangle$ and
$|1\rangle \mapsto |0\rangle$ of respective rates $\gamma_{21}$ and $\gamma_{10}$
and frequencies $\omega_{21}$ and $\omega_{10}$.
This condition is naturally found in many systems such as harmonic
oscillators undergoing dissipation, spontaneous emission of spin-1
systems or some three-level atoms.

We begin by analyzing the most favorable situation: the
complete degeneracy between the decaying
channels in each qutrit ($\omega_{21}=\omega_{10}$ and $\gamma_{21}=\gamma_{10}$), \textit{i.e.} if qutrit $A$ emits an
excitation to its reservoir, by detecting this emission there is no fundamental way to identify the corresponding decay channel (both are equally probable and generate indistinguishable excitations). 
First of all, note that in this case the no-jump evolution does not affect the entire subspace initially used by Alice and Bob,
which mimics a decoherence free subspace for this particular trajectory. 
Furthermore, contrary to the previous case, the detection of one jump
does not mean entanglement loss. For example, if Alice and Bob share the initial state $|\Psi(0)\rangle = \frac{|12\rangle+|21\rangle}{\sqrt{2}}$, 
and Alice detects an excitation in her
environment, the state of the system must still be given by $\hat{\Pi}_{1,A}|\Psi(0)\rangle$.
But now, the jump operator must include the essential fact that the
excitation does not distinguish its \textit{donor}, therefore, it
is given by $\sqrt{\gamma dt}(|1\rangle\langle 2| +
|0\rangle\langle 1|)$. When applied to $|\Psi(0)\rangle$,
this jump operator produces the state $|\Psi^{1,A}(dt)\rangle=\frac{|02\rangle+|11\rangle}{\sqrt{2}}$.
This state is still maximally entangled, when
interpreted as the state of two logical qubits. Note that the
information that the system has decayed not only allows Alice to
keep an entangled state with Bob but actually gives her the
opportunity to locally and deterministically restore $|\Psi(0)\rangle$ through the unitary feedback
of her qutrit with one excitation ($|1\rangle\mapsto
|2\rangle$, $|0\rangle\mapsto |1\rangle$). This operation should be
fast enough to avoid a consecutive decay which would surely kill
any entanglement. Since, at least in principle, this procedure could 
be repeated over and over again, entanglement could be protected for as long as necessary.
The same argument holds for Bob. In fact, the same logic can be applied to a $n$-party multi-qutrit system in which each part is subjected to dissipation but is also able to monitor the loss of excitations and to locally feedback them. Once again, as long as no part loses two excitations in a row (before the feedback mechanism comes in place), an entire subspace spanned by $\{|1\rangle,|2\rangle\}^{\otimes n}$ can be protected against local dissipation.

Another possible configuration features emitted excitations that still do not identify the
possible decay channel but are more likely the higher the number
of excitations in $|\Psi(0)\rangle$ ($\omega_{21}=\omega_{10}$ and $\gamma_{21}>\gamma_{10}$). For example, in the specific
case of dissipative harmonic oscillators, the operators in the studied 
subspace ($|n=0,1,2\rangle$) are
given by $\hat{\Pi}_0=\openone-\frac{dt \gamma}{2} (|1\rangle
\langle 1| + 2 |2\rangle \langle 2|)$ and $\hat{\Pi}_1= \sqrt{dt
\gamma} (|0\rangle \langle 1| + \sqrt{2} |1\rangle \langle 2|)$.
In this case, both possible evolutions would affect the
entanglement between Alice's and Bob's qutrits, since they
redistribute population among the levels of each subsystem. 

In Fig.~\ref{fig1}a we plot the time evolution of the entanglement of the non-monitored system, $E_F$, as well as
the average entanglement over different trajectories~\cite{AndreE} as a function of
time for the cases analyzed above, both with and without feedback.
Note that in all situations, even for the usual qubit encoding, monitoring the environments preserves
entanglement for a longer time than ignoring it, as should be the case since some information
on the system is always recovered. When the detection and feedback mechanisms are ideal ($E_{3,f}$), 
entanglement can be preserved indefinitely. In Fig.~\ref{fig1}b we plot the entanglement
obtained either when there is an $8 \%$ delay in the feedback mechanism or when there is an $8 \%$ inefficiency 
in the measurement of the environment. Since we now deal with mixed states of two qutrits, we use Negativity~\cite{Neg} to obtain the entanglement in each case. We should stress the fact that the use of qutrits is advantageous even in these non-ideal cases when compared to two qubits, as it is clear in Fig.~\ref{fig1}b.
Also note that, as expected, inefficient detection of the environment has a greater effect on entanglement
loss than delay in the feedback mechanism. 

\section{Possible implementation using Cavity Eletrodynamics}

Let us now describe a cavity QED setup  where our ideas can be applied. Suppose Alice and Bob want to establish a (ideally) perfect quantum channel. In the proposal of Ref.~\cite{Luiz}, Alice first prepares an empty cavity (state $\ket{0}$) and then interacts it with a very stable two-level atom ($\{|g\rangle,|e\rangle\}$) in the exited state $\ket{e}$, in such a way to get the entangled state $\sqrt{\alpha}\ket{0e}+\sqrt{1-\alpha}\ket{1g}$. Afterwards she sends the atom to Bob. 
Assuming that the atomic decay rate is much smaller than the cavity field one, \ie $ \gamma_{at} \ll \gamma$,
the decoherence time of the second would limit the distance achievable by the first. In fact, the
distance between Alice and Bob would have to be much smaller than $v/\gamma$, $v$ being the
atomic velocity. 

As discussed before, this strategy can be improved if Alice uses three levels of her cavity field. The idea is that, instead of being empty, the 
cavity initially contains a photon (state $\ket{1}$). Then, after the atom-cavity interaction the final state will be $\sqrt{\alpha}\ket{1e}+\sqrt{1-\alpha}\ket{2g}$. 
If no jump happens in the interval $[0,t)$ then this state evolves to
$|\tilde{\chi}_{\mathrm{NJ}}(t)\rangle=e^{-\gamma t/2}(\sqrt{\alpha}|e1\rangle+\sqrt{1-\alpha}e^{-\gamma t/2}|g2\rangle)$, with probability
$P_{\chi_0}(t)=\alpha e^{-\gamma t}+(1-\alpha)e^{-2\gamma t}$. 
However, if one jump happens in the cavity field at time $t_J<t$, then
the new state of the system at time $t$ is given by
$|\tilde{\chi}_{\mathrm{OJ}}(t)\rangle=e^{-\gamma
t_{\mathrm{J}}/2}(\sqrt{\alpha}|e0\rangle+\sqrt{1-\alpha}\sqrt{2}e^{-\gamma
t/2}|g1\rangle)$, where the no-jump operator is used for $t < t_J$ and $t > t_J$.
Note that because $\gamma_{21}=2\gamma_{10}$, the final state $|\tilde{\chi}_{\mathrm{OJ}}(t)\rangle$
is always the same irrespective of the particular time at which the jump happened. We can then treat all one-jump trajectories
as the same and compute their joint probability as $P_{\chi_1}(t)=\int_{0}^{t}P_{\chi_1}(t,t_{\mathrm{J}})dt_{\mathrm{J}}=p\left[\frac{1-e^{-\gamma t}}{\gamma}\right]$,  where $P_{\chi_1}(t,t_{\mathrm{J}})=pe^{-\gamma t_{\mathrm{J}}}$ gives the probability density for the trajectory where a jump happens at $t=t_J$ (with $p=\alpha+2(1-\alpha)e^{-\gamma t}$).
The trajectory-averaged entanglement is then given by 
$E_{2\otimes3}(\alpha,t)=E_{\chi_{NJ}}P_{\chi_0}+E_{\chi_{OJ}}P_{\chi_1}$.

\begin{figure}[h!]
\includegraphics[width=8.5cm]{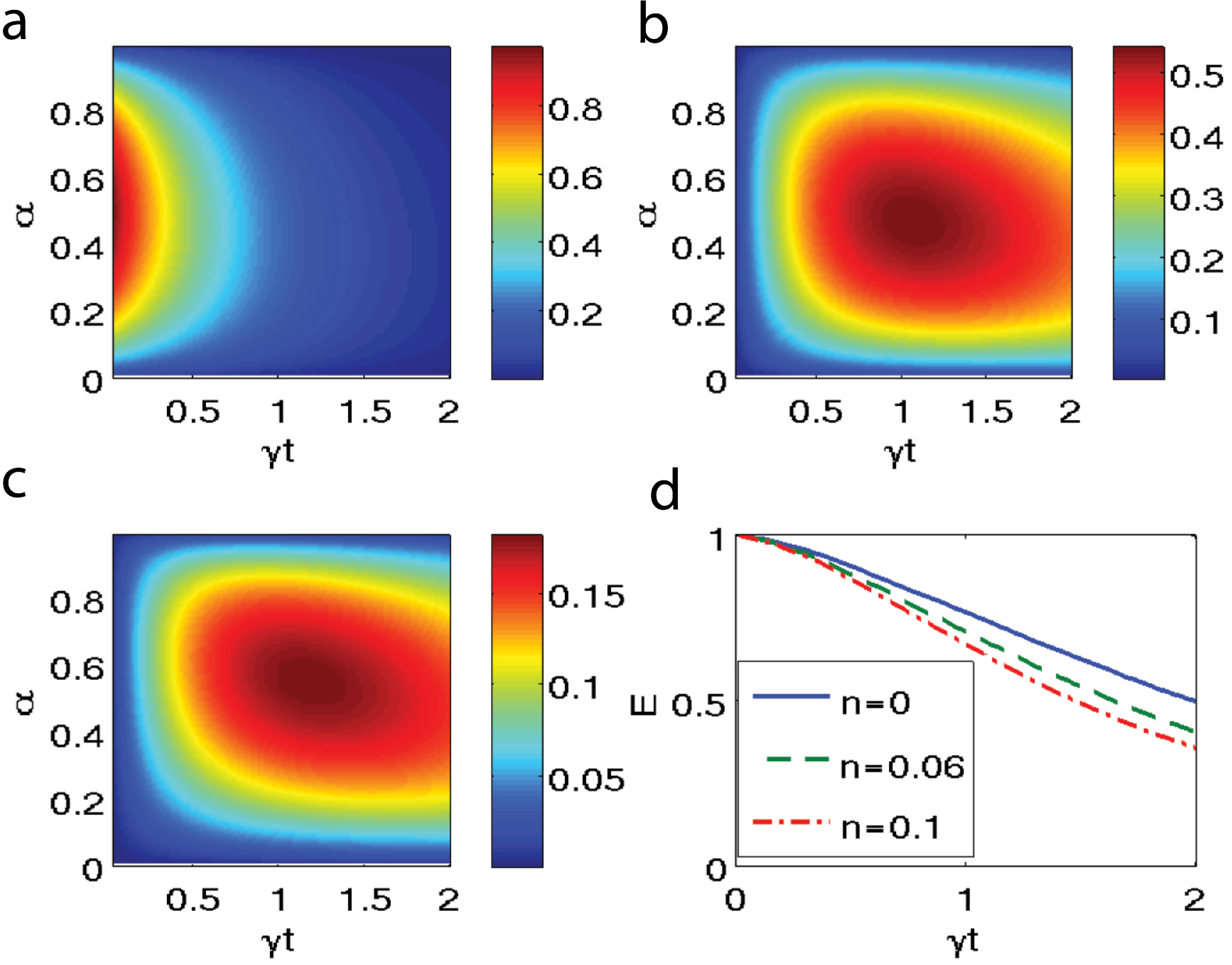}\\
  \caption{(Color Online) a. No-jump contribution $E_{\chi_{NJ}}P_{\chi_0}$ for $E_{2\otimes 3}$; b. (upper right) one-jump contribution $E_{\chi_{OJ}}P_{\chi_1}$ for $E_{2 \otimes 3}$; c.  $E_{2\otimes3}-E_{2\otimes2}$; d. $E_{2\otimes3}$ with $\alpha=0.5$ for different temperatures with the mean environment excitation number given by $n$. Decay time $1/\gamma=0.129s$ as in the experimental setup in~\cite{HarocheNature}.} 
\label{Superficie2}
\end{figure}

Furthermore, we can choose $t$ to maximize the one-jump trajectory and $\alpha$ to maximize the entanglement
of the state when the atom reaches Bob, hence, creating an even longer lasting entanglement. 
Fig.~\ref{Superficie2} compares the protocols as a function of $t$ and $\alpha$
showing the optimum range in which each strategy is better and also that
for all times and initial states, encoding qubits in qutrits is more efficient in the
presence of decay and environmental monitoring. Fig.~\ref{Superficie2}d shows
entanglement for different temperatures of the reservoir. Typical microwave Cavity QED experiments
correspond to the intermediate curve~\cite{HarocheNature}. Note that at such low temperatures, thermal excitations play a
minor role in the proposed scheme. 

\section{Applications for communication protocols}

The schemes proposed before can be
readily incorporated in some previously known communication
protocols. One of the main ideas of transmitting quantum
information in 1-D networks relies on quantum repeaters
\cite{Repeat}. In this case, one wants to transmit entanglement
through large distances and uses intermediate stations to recover
it from time to time through entanglement distillation. Once some
degree of entanglement is recovered in an intermediate station,
this entanglement is teleported to the next station. Naturally,
since using qutrits allows to keep entanglement for longer times
between each station, there is a reduction in the total amount of
necessary resources.

In higher dimensional networks, it was recently shown that the
geometry of the graph defining the quantum network plays an
important role in the problem of long-distance communication. Entanglement Percolation ideas \cite{EntPerc} were first
presented for pure-state based networks.  The main building block for these strategies is the possibility of performing the optimal singlet conversion. As we discussed before, this task can be achieved by reservoir monitoring. This fact in turn shows a strategy of realizing Classical Entanglement Percolation in noisy lattices \cite{MixedPerc}.


\section{conclusion}

To conclude, we have shown that for systems undergoing
dissipation, encoding qubits in qutrits preserves coherence and
entanglement for much longer times if the excitations given to the
local reservoirs can be detected by external observers. Thus,
systems with their environments under continuous measurement are
not only suitable but advantageous for some quantum information
protocols. Moreover, the monitoring scheme is further
improved if the parties are able to locally feedback the recovered information into the system. This leads naturally to a reduction in resources 
for protocols such as quantum repeaters, teleportation, swapping and error correction, as well as 
an improvement in the coherence time of quantum memories. We
have also shown a way of using the information leakage to the environment to perform the optimal singlet conversion, the central task in Classical Entanglement Percolation. Finally, note that experiments already observe quantum jumps in different systems, e.g, harmonic oscillators (microwave cavity fields)~\cite{HarocheNature}, and in single ions~\cite{ion}, which means that the scheme here proposed is clearly within nowadays technology.

\begin{acknowledgements}

The authors thank A.~Ac\'in and A.R.R.~Carvalho for enlightening
discussions. Support from Brazilian agencies CNPq and Fapemig and
the European project QAP is warmly acknowledged. This work is part
of Brazilian National Institute of Science and Technology on
Quantum Information.

\end{acknowledgements}

\end{document}